%% file: main.tex
\documentclass[acmtog, nonacm]{acmart}
\acmSubmissionID{1152}

\usepackage{booktabs} 
\usepackage{enumitem}
\usepackage{amsmath}
\usepackage{bm}
\usepackage{svg}
\usepackage{sidecap}
\usepackage{pifont}
\input{commands}

\usepackage[ruled]{algorithm2e} 

\SetAlFnt{\small}
\SetAlCapFnt{\small}
\SetAlCapNameFnt{\small}
\SetAlCapHSkip{0pt}

\acmJournal{TOG}

\begin{document}

\title{Neuralocks: Real-Time Dynamic Neural Hair Simulation}

\author{Gene Wei-Chin Lin}
\affiliation{%
  \institution{Meta Reality Labs}
  \country{Canada}
}

\author{Egor Larionov}
\email{elrnv@meta.com}
\affiliation{%
  \institution{Meta Reality Labs}
  \country{USA}
}

\author{Hsiao-yu Chen}
\email{hsiaoyu@meta.com}
\affiliation{%
  \institution{Meta Reality Labs}
  \country{USA}
}

\author{Doug Roble}
\email{droble@meta.com}
\affiliation{%
  \institution{Meta Reality Labs}
  \country{USA}
}

\author{Tuur Stuyck}
\email{tuur@meta.com}
\affiliation{%
  \institution{Meta Reality Labs}
  \country{USA}
}
\renewcommand{\shortauthors}{Lin et al.}

\begin{abstract}

Real-time hair simulation is a vital component in creating believable virtual avatars, as it provides a sense of immersion and authenticity. The dynamic behavior of hair, such as bouncing or swaying in response to character movements like jumping or walking, plays a significant role in enhancing the overall realism and engagement of virtual experiences. Current methods for simulating hair have been constrained by two primary approaches: highly optimized physics-based systems and neural methods. However, state-of-the-art neural techniques have been limited to quasi-static solutions, failing to capture the dynamic behavior of hair. This paper introduces a novel neural method that breaks through these limitations, achieving efficient and stable dynamic hair simulation while outperforming existing approaches. We propose a fully self-supervised method which can be trained without any manual intervention or artist generated training data allowing the method to be integrated with hair reconstruction methods to enable automatic end-to-end methods for avatar reconstruction. Our approach harnesses the power of compact, memory-efficient neural networks to simulate hair at the strand level, allowing for the simulation of diverse hairstyles without excessive computational resources or memory requirements. We validate the effectiveness of our method through a variety of hairstyle examples, showcasing its potential for real-world applications.

\end{abstract}


\begin{teaserfigure}
  \includegraphics[width=\textwidth]{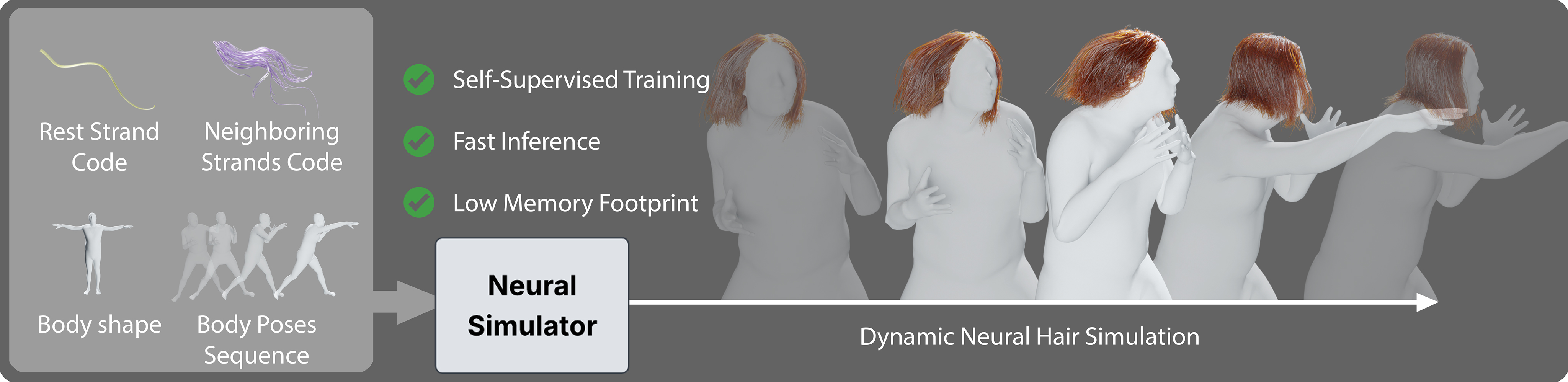}
  \caption{We present Neuralocks, a novel method that achieves high-performance dynamic neural hair simulation with a compact network size. This makes it a suitable solution for deployment on lower-end devices, which are commonly used in games and interactive experiences. At the core of Neuralocks lies its ability to generate strand-level deformations by leveraging boundary condition history and local lock neighborhood information, enabling realistic and detailed hair simulations. Our method is fully self-supervised, eliminating the labor intensive and expensive process of generating simulated training data.} 
  \label{fig:teaser}
\end{teaserfigure}

\maketitle

\section{Introduction}

Creating believable virtual characters is a vital aspect of visual effects, gaming, and immersive experiences. One crucial element that significantly enhances the authenticity of these digital representations is hair. The realistic motion of hair plays a pivotal role in generating engaging and convincing virtual environments. In particular, the dynamic behavior of hair in response to body movement is essential for creating a sense of realism and immersion. Although commercial digital content creation tools have made significant strides in simulating realistic hair motion, particularly in offline applications for high-end visual effects, the transition to real-time interactive environments remains a formidable challenge. The computational demands of simulating believable hair are substantial, and when combined with the need to share resources with other critical components such as gameplay mechanics or rendering, achieving real-time performance becomes increasingly difficult. 

Despite the challenges posed by computational requirements, researchers have made notable strides in exploiting GPU parallel compute capabilities to produce real-time hair simulations~\cite{daviet2023interactive}. However, the reliance on specialized hardware remains a limitation, making it difficult to achieve similar frame rates on commodity hardware or mobile devices. Recently, state-of-the-art methods such as GroomGen~\cite{zhou2023groomgen} and Quaffure~\cite{stuyck2024quaffure} have addressed this issue by leveraging the efficiency of neural networks.

In addition to performance considerations, it is crucial to develop methods that are fully automatic to facilitate seamless character generation. This field has seen a significant surge in interest with applications in realistic avatar telepresence using virtual reality headsets~\cite{saito2024rgca, donglai, ma2021pixel}, where researchers strive to create realistic avatars from images and videos in a completely automated manner. Hair reconstruction is an active area of research in this domain~\cite{sklyarova2023neural, zakharov2024human} but these methods primarily focus on strand reconstruction and omit producing a deformation model. GroomGen proposes a data-driven method for modeling deformations which relies on artist generated training data, thereby creating a major manual bottleneck in automating the learning process for obtaining a generalizeble simulation model for hair. Quaffure addresses this limitation by introducing a self-supervised training setting for learning hair deformations based on body pose and shape. Nevertheless, both these solutions are still limited to quasi-static drapes of hair.

To address this challenge, we propose a fully automatic, self-supervised method that can be trained without manual intervention or artist-generated training data. This approach can be seamlessly integrated with existing hair reconstruction techniques, enabling the development of end-to-end systems for automatic character generation, which runs at high frame rates. Our proposed model demonstrates better generalization, runtime performance as well as a much lower memory footprint compared to Quaffure. We are the first to produce dynamic neural hair simulations leveraging a completely self-supervised training procedure.

Although prior work on the self-supervised learning of dynamic cloth exists~\cite{bertiche2022neural, santesteban2022snug}, our proposed method effectively learns a mapping from body pose history to dynamic hair results without the need for modeling time evolution processes. This greatly simplifies the training and inference of the networks by eliminating the need for Gated Recurrent Units (GRUs)~\cite{bertiche2022neural, santesteban2022snug} and hidden states, and enables our method to start at any point in the animation.

We overcome key limitations of previous methods GroomGen and Quaffure whilst improving performance and quality. Both methods are restricted to quasi-static deformations. GroomGen relies on a data-driven approach that would require the generation of an extensive dataset to capture dynamics — a labor-intensive process, which is impractical at scale. Quaffure circumvents the need for large datasets through a self-supervised framework, but its formulation still omits dynamic behavior and  because it predicts the full groom directly, their approach results in a network architecture that is prohibitively large for deployment on mobile devices. In contrast, we introduce a fully self-supervised approach capable of efficient neural hair dynamics using neural networks that are orders of magnitude more compact than Quaffure, making them suitable for resource-constrained environments. In summary, our contributions are :
\begin{itemize}
    \item the first self-supervised method for dynamic neural hair simulation, eliminating the need for manual intervention or training data generation. This approach enables deterministic, stable, and dynamic simulation results. Building upon our novel hair solver, we further contribute a self-supervised extension to Quaffure, empowering it to handle dynamic deformations.
    \item an improved Cosserat formulation for hair strand modeling, achieving a substantial enhancement in quality compared to previous approaches while preserving competitive runtime performance.
    \item a novel hair style-preservation losses that effectively maintain hairstyle integrity under dynamic motion, and introduce a proximity-based augmentation technique for the network input, leading to improved results.
\end{itemize}
We showcase the effectiveness of our method across a diverse range of hair styles and body shapes, highlighting its adaptability for both strand-based and mesh-based rigged hair.

\section{Related Work}

\paragraph{Physics-based Simulation} Simulating hair has garnered significant attention for decades, with earliest works dating back to the 90s \cite{anjyo1992simple, daldegan1993integrated}, which simulated cantilever beam equations per strand providing the ability to model hair in a physically plausible way. Later, researchers have developed new ways to represent hair in a simulation using volumes \cite{hadap2001modeling}, super-helices~\cite{bertails2006}, Cosserat rods~\cite{pai2002strands, sca2016cosserat}, spring networks \cite{selle2008}, and meshes \cite{yuksel2009hair}. Each representations carries a certain trade-off between quality, physical accuracy, performance and robustness. In all these cases, however, hair contact has been a notoriously difficult problem, and has been motivated a major focus in the years to follow \cite{choe2005haircontact, daviet2011, bertails2011}.  We follow the work of \citet{sca2016cosserat} for representing hair physics and \citet{hadap2001modeling} for addressing collisions between hair strands by treating it as a volume. This provides a simple, efficient, and differentiable foundation, which fits well with our neural approach. Later works also corrected artificial sagging \cite{derouet2013inverse, twigg2011sagfree, hsu2023sag} to improve modeling accuracy, which explored using high order derivatives to solve the inverse simulation problem, but required specialized solvers. \citet{guan2012multi} proposes an efficient data-driven approach for learning hair models to enable efficient runtime performance. This marks some of the earliest work of using hair simulation as a precomputation to improve simulated results. This idea is paramount to the neural approach we follow for simulating hair, and has been widely explored. In recent years, researchers have also targeted the computational complexity of simulating hair by developing new algorithms for simulating hair on the GPU \cite{daviet2023interactive, edwin} and further extended hair simulation scope to highly curly or kinky hair \cite{WUSHI2023, HWU2024} as well as efficient hair interpolation \cite{hsu2024real}. Most recently, \citet{Hsu2025} presented a novel solver that achieves high accuracy even at high stiffness levels, while maintaining stability under large time steps. To this day, using physics at run-time to simulate hair remains a luxury since compute budget for animation is typically constrained on mobile or mixed reality devices.

\paragraph{Self-supervised Learning} A recent trend in learning elastic deformation of objects has leveraged self-supervised learning to overcome the limitations of supervised methods. By eliminating the need for large, labeled datasets, self-supervised learning reduces data requirements and improves transferability due to its agnosticism towards material, shape, and collisions. This approach has shown successful results in learning cloth deformation driven by articulated characters, both in quasi-static~\cite{de2023drapenet} and dynamic scenarios~\cite{santesteban2022snug, grigorev2023hood, bertiche2022neural}.
Self-supervised learning has also been applied to learn quasi-static deformation of hair, enabling realistic simulations of different hairstyles and poses~\cite{stuyck2024quaffure}. Additionally, it has been used to learn reduced ordered kinematics for rigid and elastic objects~\cite{sharp2023data}, with performance improvements achieved through the enforcement of Lipschitz regularization~\cite{lyu2024accelerate}. Furthermore, subspace neural simulation work has leveraged self-supervised learning to enhance motion with nonlinear modes~\cite{wang2024neural}. 

\paragraph{Neural Hair Simulation and Reconstruction} Neural methods for hair simulation have been gaining attention, with early work by \citet{lyu2020real} presenting a neural interpolator for hair upsampling based on a CNN framework. \citet{chen2024doubly} propose a hierarchical generative representation for strand-based hair geometry to capture high-frequency detail. Perm~\cite{he2024perm} presents a learned parametric representation by decomposing strands in frequency space. Haar~\cite{sklyarova2023haar} presents a text-conditioned generative model for hair strands. GroomGen~\cite{zhou2023groomgen} introduced a generative hair model for static grooms and a data-driven quasi-static simulator for hair, which relies on simulated hair data for training.
Recently, a large collection of work has focused on applying neural networks to recover hair geometry and appearance from sparse observations such as images, videos or 3D scans. Several works consider the controlled capture setting which leverages multiview dome captures~\cite{wang2022neuwigs, Wang_2022_CVPR, wang2024local}. Gaussian Haircut~\cite{zakharov2024human} leverages a dual strand and Gaussian representation to enable hair strand and appearance reconstruction from multi-view information. Most recently, DiffLocks~\cite{difflocks2025} leverages a large synthetic dataset to construct a diffusion model that is capable of predicting hair grooms from a single image only. \citet{Chang2025IPHG} presents a method to transform unstructured reconstructed hair strands into procedural grooms to enable artist-friendly editing.

\section{Method}

\begin{figure*}
    \centering
    \includegraphics[width=0.85\linewidth]{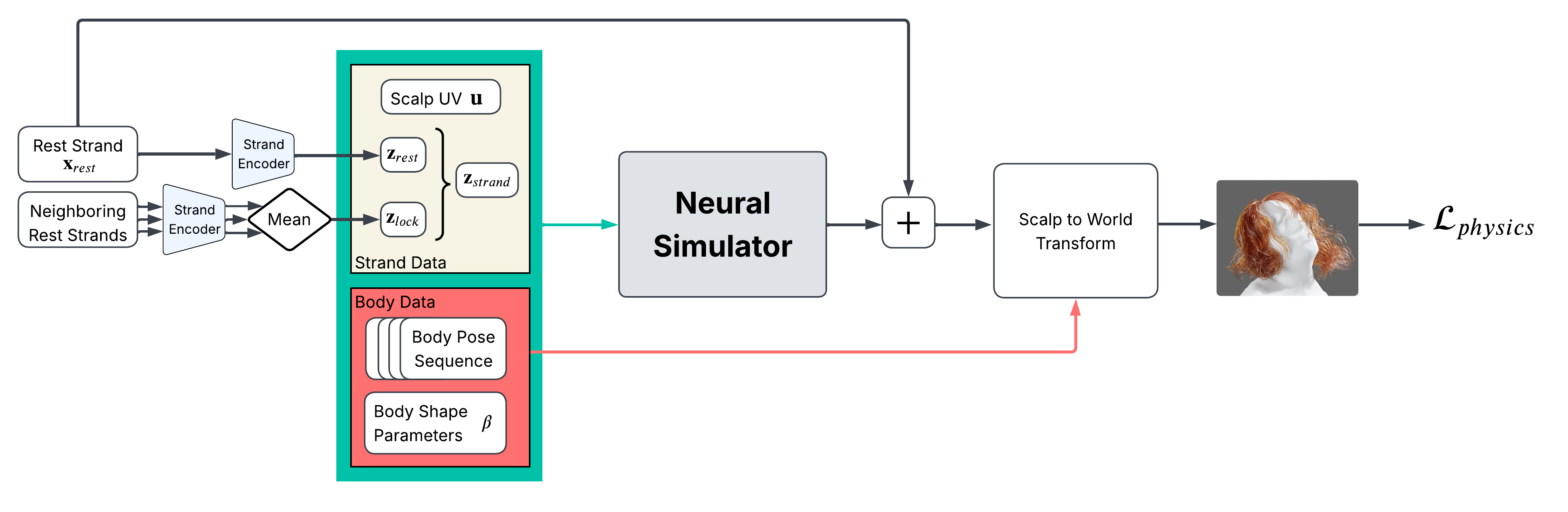}
    \caption{\textbf{Method Overview:} Our method achieves high-quality real-time neural hair simulation by leveraging a two-stage training procedure. (i) We train a strand encoder to encode the full space representation into a compact latent code, given the rest shape of the grooms. (ii) The second stage trains the dynamic neural strand simulator to produce dynamic results conditioned on body shape and pose history, as well as the rest strand latent encoding and and encoding of its local strand neighborhood. The neural simulator predicts the local displacements in a canonical representation that gets positioned in world space using a rigid based on body pose and shape.}
    \label{fig:pipeline}
\end{figure*}

We present an efficient method for predicting the dynamic 3D deformation of hair under various body movements and shapes, with limited memory and compute requirements. We adopt a lightweight strand-based neural model that predicts the deformation of a single strand at a time.
We achieve this through a two-step training procedure. First, we train an autoencoder to compress our strand representation into a latent representation to improve efficiency and memory requirements. We keep the encoder and freeze the weights and discard the decoder. With the fixed strand encoder, we train our dynamic neural strand-based simulator in a fully self-supervised approach without the need for simulated hair as training data.
Our method overview is shown in Figure~\ref{fig:pipeline}.

\subsection{Preliminaries}

\subsubsection*{Body and Motion}

Our work leverages a statistical body model to produce posed bodies with a variety of parameterized body shapes. We obtain body motion sequences from in-house motion capture recordings. 

\subsubsection*{Grooms}

Our results are produced by leveraging the CT2Hair groom dataset~\cite{shen2023CT2Hair}. In addition, we complement our groom library with a variety of artist-made grooms. Note that our method is not limited to this dataset and other strand-based grooms can be used for training. 

\subsection{Strand Encoder}

We employ a strand encoder to reduce the full space strand shape encoding into a compressed representation. Each strand vertex is encoded relative to the root position $\textbf{x}_0$ in a local scalp space $\textbf{x}_i = \textbf{T}(\textbf{x}_i - \textbf{x}_0)$, where $\textbf{T}$ is the transformation from the world space to the scalp space. Although prior work proposes a more complex strand encoders design \cite{zhou2023groomgen}, we design our method with performance in mind and found that a standard 2-layer MLP with 256 hidden units is sufficient to reconstruct the wide variety of hairstyles used in all of our examples. However, more complex strand encoders are equally compatible with our subsequent pipeline. In our implementation, the latent space representation $\textbf{z}$ is encoded in $\mathbb{R}^{32}$. We train the strand encoder on all strands in our static groom library.

\subsection{Neural Strand Simulator}

At the core of our method is a neural strand simulator which is designed with efficiency, stability and generalizability in mind. The simulator maps a history of boundary conditions and a latent description of the strand and its proximity-based lock information directly to a deformed state without the need of modeling a time evolution process for the hair strands themselves.  

To distinguish individual strands, our input consists of the UV coordinate embedding $\textbf{u}$ of the strand root vertex on the scalp, rest strand latent vector $\textbf{z}_{strand} = \{\textbf{z}_{rest}, \textbf{z}_{lock}\}$, where $\textbf{z}_{rest}$ is the rest strand latent code obtained from the strand encoder and $\textbf{z}_{lock}$ is the average of the neighboring strand latent code, which is used to further distinguish strands across multiple grooms. The number of neighboring strands is determined by a clustering process based on the distribution of the strand roots and the size of the cluster. To inform the network of the body configuration, we concatenate the current body pose and shape coefficients. Finally, to enable the neural simulator to model dynamic results, we include a concatenation of the previous N body joint velocities. The output of the neural strand simulator are the displacement vectors which are added to the embedding of the strands in canonical space which is then rigidly transformed to obtain the vertices of the deformed strand in world space. In contrast to the world space offset prediction in Quaffure, our canonical space effectively filters out character rotations which simplifies the task of the network.

\subsection{Physics-based Loss Formulation}

We build upon the Cosserat rod model~\cite{sca2016cosserat}, adopting the formulation for stretch, gravity, and collision potentials as presented in Quaffure~\cite{stuyck2024quaffure}. Notably, we introduce a novel bend-twist model that surpasses previous approaches, yielding improved performance and results. Furthermore, we incorporate an inertia potential to capture hair dynamics, drawing inspiration from~\citet{santesteban2022snug}. However, we modify the boundary condition history conditioning to eliminate the need for modeling a time evolution process, thereby obviating the requirement for GRU models. This simplification significantly streamlines training and inference. The introduction of the inertial potential $\mathcal{L}_\text{inertia}$ renders a pose regularization terms unnecessary. Additionally, in contrast to prior work, we introduce novel loss formulations to maintain hairstyle and clumping behavior to produce more realistic results and to allow for artistic direction when desired. Our total loss is the sum of all listed losses.
\begin{equation}
\begin{aligned}
 \mathcal{L}_\text{physics} &= \mathcal{L}_\text{inertia} + \mathcal{L}_\text{gravity} + \mathcal{L}_\text{elastic} + \mathcal{L}_\text{collision} + \mathcal{L}_\text{hair},
\end{aligned}
\end{equation}
where $\mathcal{L}_\text{elastic}$ consists of the rod-specific elastic potentials $\mathcal{L}_\text{stretch}$ and $\mathcal{L}_\text{bend\_twist}$, $\mathcal{L}_\text{collision}$ includes strand-body collision potential $\mathcal{L}_\text{body\_collision}$ and strand self-collision $\mathcal{L}_\text{self\_collision}$. We also include hair-specific potential $\mathcal{L}_\text{hair} = \mathcal{L}_\text{hair\_style} + \mathcal{L}_\text{adhesion}$ for constrained hairstyles and hair-hair interaction, respectively.

\subsubsection{Elastic Potentials}

A Hookean potential models strand inextensibility by penalizing the difference between the length $l$ of each hair segment and its rest length $l_{rest}$, evaluated from the initial strand configuration:
\begin{equation}
\mathcal{L}_\text{stretch} = \frac{\texttt{k}_\text{stretch}}{2} \sum_\text{edges}  \left(l - l_{rest}\right)^{2}.
\end{equation}

\citet{stuyck2024quaffure} proposed a modified Cosserat potential to reduce the impractical training time from naively using the full Cosserat model. While the modified Cosserat potential effectively produces plausible deformation, the range of deformation is limited due to the constant unit director of the edge used to evaluate the potential, resulting in overly stiff hair animations. Instead, we propose an improved Cosserat potential that allows for a wider range of deformations, producing more plausible strand shapes while maintaining similar performance characteristics. Specifically, our modified bend-twist model is in the form:
\begin{equation}
\mathcal{L}_\text{bend\_twist} = \frac{\texttt{k}_\text{bend\_twist}}{2} \sum_\text{edge pairs} l_{rest} \mathbf{\Omega}^\top\mathbf{\Omega},
\end{equation}
where $\mathbf{\Omega} = \frac{2}{l_{rest}}\left(\Im\left(\bar{q}_iq_{i+1}\right) - \Im\left(\bar{q}_i^0q_{i+1}^0\right)\right)$ is the discrete bend-twist strain measure which requires evaluating each consecutive hair orientation pair $q_i$ and $q_{i+1}$, with rest orientations $q_{i}^0$ and $q_{i+1}^0$. In our implementation, we evaluate the hair orientations $q_i$ by parallel transporting from the root segment $e_0$ to the current segment $e_i$. This greatly improves the performance comparing with computing parallel transport from the root segment until the current segment, which was identified as the computational bottleneck in previous work~\cite{stuyck2024quaffure}. 

\subsubsection{Collision Potentials}

To model the collision potential and the collisions between hair strands and the body mesh as well as the hair collisions with itself, we adopt the collision models presented by \citet{stuyck2024quaffure}. Specifically, the body collision model maintains a minimal distance $D$ between a predicted hair vertex and the outward facing normal of the body geometry. 
\begin{equation}
\mathcal{L}_\text{body\_collision} =  \texttt{k}_\text{body\_collision} \sum_\text{vertices} \text{max} \left( D - d \left( \mathbf{x} \right), 0 \right)^3,
\end{equation}
where $d$ computes the signed distance along the body triangle normal direction and $\texttt{k}_\text{bc}$ is the collision stiffness. Self-collisions are modeled using SPH density estimation 
\begin{equation}
\begin{aligned}
\mathcal{L}_\text{self\_collision} &= \texttt{k}_\text{sc} \sum_\text{vertices} \text{max} \left(\rho(\mathbf{x}) - \rho_\text{rest}, 0 \right)^3, \\
\rho(\mathbf{x}) &= \sum_j m_j W \left( ||\mathbf{x} - \mathbf{x}_j||, h \right),
\end{aligned}
\end{equation}
where $W\left(||\mathbf{x} - \mathbf{x}_j||, h\right)$ is the smoothing kernel proposed by \citet{bando2003animating}.

\subsubsection{Gravity}

The gravity potential with hair positions $\mathbf{x}$ is modeled as:
\begin{equation}
\mathcal{L}_\text{gravity} = \sum_\text{vertices} -m \mathbf{g}^\top \mathbf{x},
\end{equation}
where $\mathbf{g}$ is the gravitational acceleration and $m$ is the particle point mass.

\subsubsection{Style Preservation}

\begin{figure}
    \centering
    \includegraphics[width=0.95\linewidth]{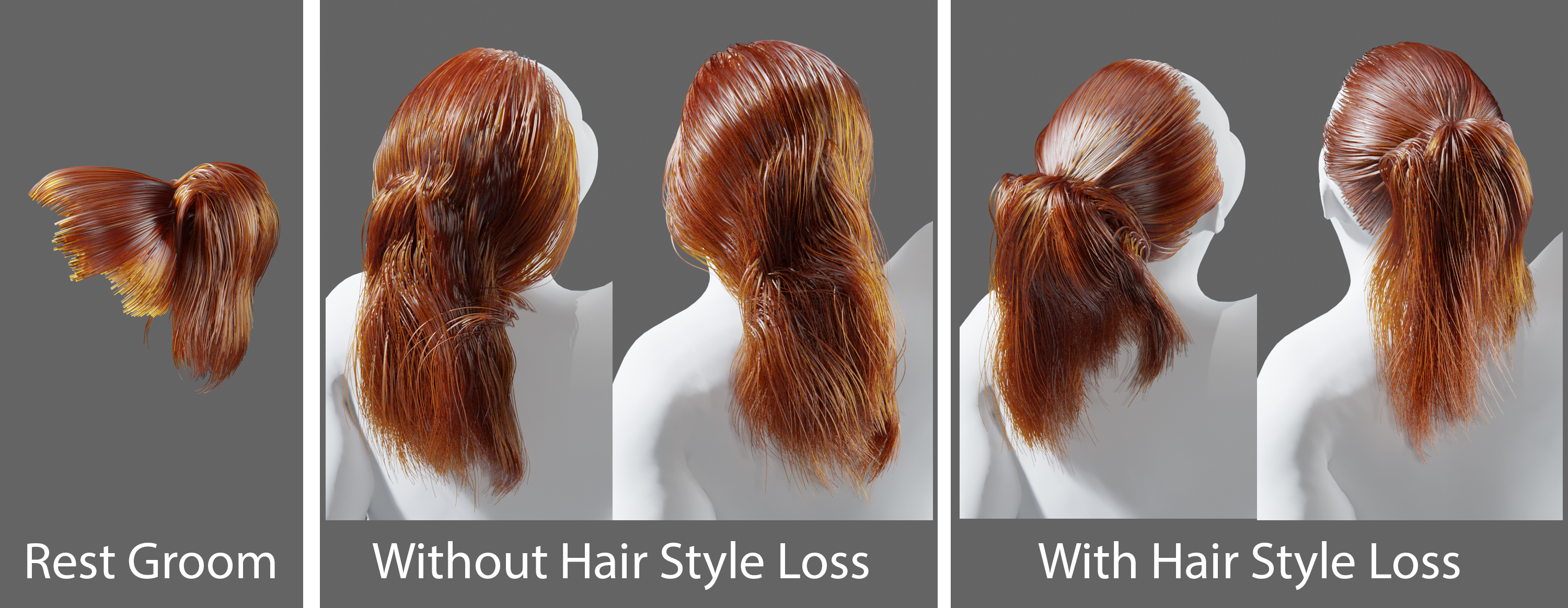}
    \caption{\textbf{Hair Style Loss Comparison} Saggy strands can be observed without our proposed loss (middle), while the ponytail shape of the constrained parts of the strands is well preserved throughout the animation when the hair style loss is added (right).}
    \label{fig:hairstyleLossComparison}
\end{figure}

We introduce an additional hair style loss for some hairstyles with accessories such as hair ties to constrain the strands to maintain the style throughout the simulation:
\begin{equation}
\begin{aligned}
 \mathcal{L}_\text{hair\_style} &= \texttt{k}_\text{hair\_style}\sum_\text{vertices} \left( 1 - min\left(1, \frac{s \left(\mathbf{x}\right)}{s_{max}} \right) \right)\left(\mathbf{x} - \mathbf{x}_{posed}\right)^{2}
 \end{aligned}
\end{equation}
where $s$ computes the closest distance to the body mesh and $s_{max}$ is a user selected style-specific threshold. Fig.~\ref{fig:hairstyleLossComparison} shows a significant quality improvement with the proposed hair style loss. 

\begin{figure}
    \centering
    \includegraphics[width=0.95\linewidth]{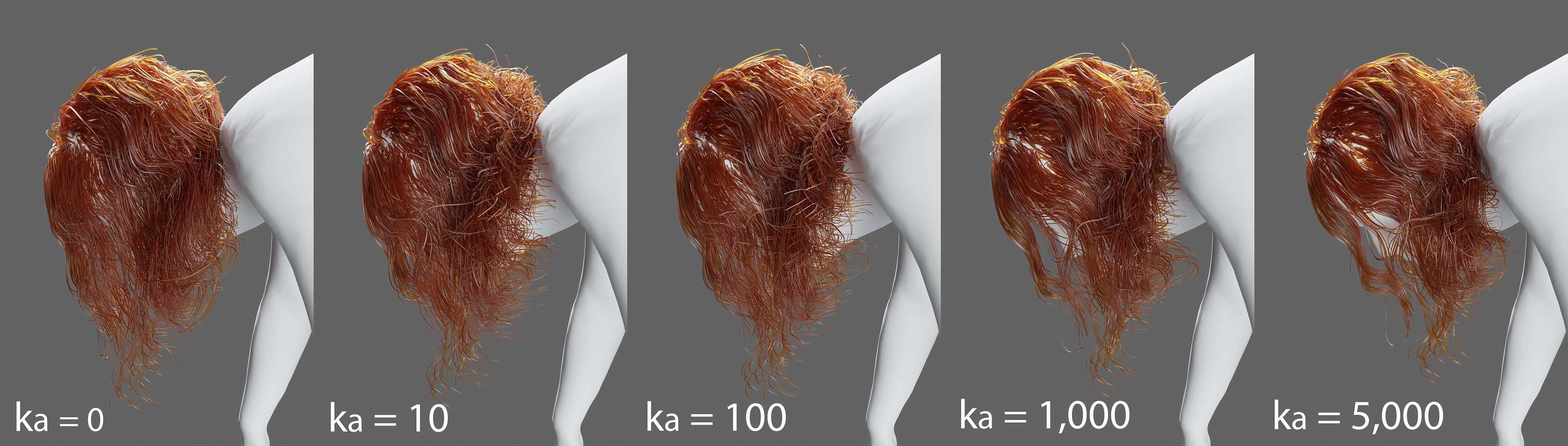}
    \caption{\textbf{Adhesion loss} Our adhesion loss effectively captures the internal friction and static electricity between strands, allowing for a more realistic simulation of hair behavior. The figure below illustrates the impact of varying adhesion stiffness on hair dynamics. Left to right, we observe how increasing adhesion stiffness causes the hair to behave more like clumps, with individual strands becoming increasingly influenced by their neighbors.}
    \label{fig:adhesion}
\end{figure}

Additionally, we model the effect caused by internal friction and static electricity between strands with the adhesion loss:
\begin{equation}
\begin{aligned}
 \mathcal{L}_\text{adhesion} &= \texttt{k}_\text{adhesion}\sum_\text{vertices}\left(\mathbf{x} - \mathbf{x}_{neighbor}\right)^{2},
\end{aligned}
\end{equation}
where we compute five closest neighboring vertices $\mathbf{x}_{neighbor}$ of a hair vertex $\mathbf{x}$ within a predefined distance $\mathbf{r}_{neighbor}$ at the rest pose. We set $\mathbf{r}_{neighbor}$ to 0.25 in all of our examples. The adhesion loss effectively preserves clumps existed in the original hairstyle during the simulation as shown in Fig.~\ref{fig:adhesion}.

\subsubsection{Dynamics}
\label{subsec:dynamics}

The losses defined in the previous subsection are sufficient to train a quasi-static simulator. In order to obtain dynamic results, we include an additional inertia term to our total energy~\cite{bertiche2022neural, santesteban2022snug} which is derived from the variational formulation of Newton's method, recasting it as an optimization problem~\cite{gast2015optimization}. During training, we let the network predict three consecutive frames. The first two are used to compute the inertial prediction $\hat{\mathbf{x}} = 2 \mathbf{x}_{t-1} - \mathbf{x}_{t-2}$ and are detached from the computation graph to prevent back propagation into past states. We then compute an inertia loss on the last predicted frame $\mathbf{x}$ as 
\begin{equation}
\begin{aligned}
 \mathcal{L}_\text{inertia} &= \frac{1}{ 2 \Delta t^2} \left(\mathbf{x} - \hat{\mathbf{x}} \right)^\top \mathbf{M} \left(\mathbf{x} - \hat{\mathbf{x}} \right)
\end{aligned}
\end{equation}
What sets our approach apart is the lack of recurrent training component which greatly simplifies neural network training and inference as the network is entirely conditioned on the sequence of input boundary conditions and not the current state of the hair. Prior works use GRUs~\cite{santesteban2022snug, bertiche2022neural} which require hidden state initialization and additional considerations during training. In contrast, our approach provides a deterministic mapping with stable results which greatly simplifies training and inference as it allows us to 
randomly select a pose with boundary condition history and body shape parameters for each training step. We then evaluate our self-supervised losses without the need for simulated hair data.
 
\section{Results}

We train our neural simulator on all strands from 10 selected grooms at the same time. During training, we randomly select one frame from a motion in our motion capture dataset, along with pose history $N$ and a random body shape. We use pose history $N = 30$ in all dynamic results. Once trained, our neural network generates plausible strand deformation corresponding to the body motions. Unless otherwise stated, all results contain grooms from the training set but demonstrate body motion and shape generalization. In addition, our neural simulator is able to generalize to similar grooms that are not in the training set as shown in Fig.~\ref{fig:unseenGrooms}.

\subsection{Comparison to Related Work}

We provide comparisons to recent state-of-the-art works and demonstrate that our model is the only one capable of producing dynamic hair deformations within a small compute and memory footprint.

\subsubsection*{GroomGen}

\begin{figure}
    \centering
    \includegraphics[width=0.95\linewidth]{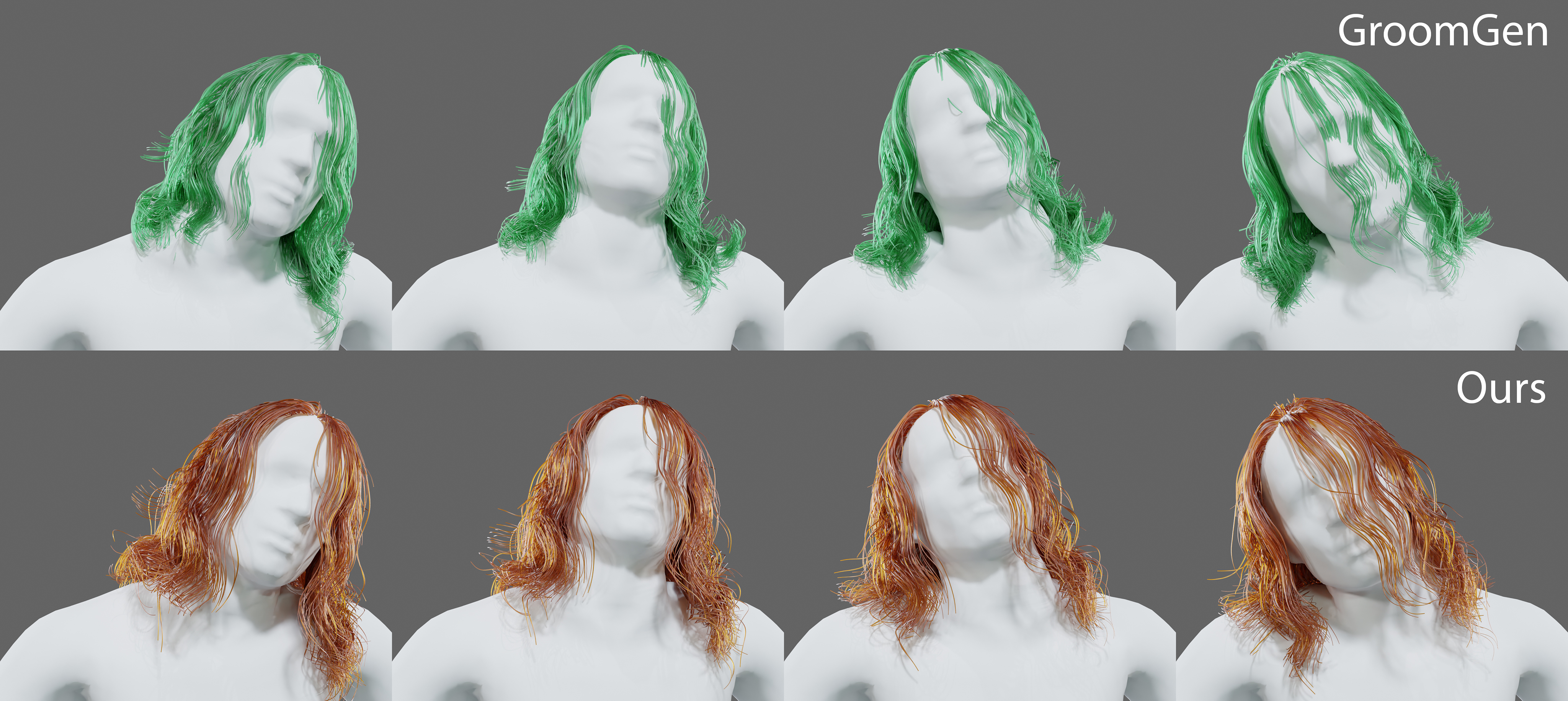}
    \caption{\textbf{GroomGen Comparison} Our method produces dynamic simulation results without the need for simulated training data, whereas GroomGen is quasi-static and requires pre-simulated data. Our method also produces higher visual quality results with noticeably fewer intersections of the hair strands with the body as can be seen by the frequent intersections of the strands with the nose in the case of GroomGen.}
    \label{fig:groomgenComparison}
\end{figure}

\citet{zhou2023groomgen} presents a neural quasi-static simulator that takes a strand-based simulation approach which is trained in a data-driven manner using precomputed simulation data. Fig.~\ref{fig:groomgenComparison} shows that our model is able to produce higher quality dynamic results with much fewer intersections with the body and is able to do so at higher frame rates as discussed in Section~\ref{subsec:performance}. 

\subsubsection*{Dynamic Quaffure}

\citet{stuyck2024quaffure} predicts hair deformation for the entire groom in a single forward pass.
The model is limited to quasi-static deformations, to provide a fair comparison, we extend the method to include dynamics using the procedure described in Sec.~\ref{subsec:dynamics}, maintaining the self-supervised training procedure. We leverage the same representation and network structure but augment the network input with the joint angle velocities of the previous 30 frames. With this increased input size, the model size amounts to a total of 692MB. In comparison, our dynamic model weighs in at only 1MB. Fig.~\ref{fig:quaffureComparison} shows that both approaches produce comparable results but ours does so at a significantly lower memory and performance footprint as discussed in Section~\ref{subsec:performance}. In addition, our Cosserat formulation reduces the need for parameter tuning to obtain the desired results. Quaffure predicts the full groom in a single forward pass which limits the maximum strand count to 4096 in their implementation whereas ours scales gracefully with strand count. Increasing the number of strands for Quaffure necessitates larger maps, which results in increases in both memory usage and runtime.

\subsubsection*{Physics-based Simulation}

Simulation has long been the gold standard for achieving high-quality and real-time results. As a benchmark, we compare our approach to a high-performance XPBD simulation model~\cite{macklin2016xpbd}, which is widely employed in real-time applications due to its ability to leverage GPU hardware capabilities.  Our proposed network offers a significant advantage by offloading a substantial amount of computations to training time, resulting in orders of magnitude faster inference, as detailed in Table.~\ref{tab:runtime_performance_table}. Fig.~\ref{fig:simVSNeuralocks} illustrates that our method produces reasonable results that capture the overall dynamics comparable to those obtained by physics-based simulation, albeit with a more damped motion. The reduction in quality is expected considering that our method is about 740 times faster.

\subsection{Ablation Studies}

We ablate several key components of our method to demonstrate its contributions to producing high quality results.

\subsubsection*{Lock Based Dynamics}

\begin{figure}
    \centering
    \includegraphics[width=0.95\linewidth]{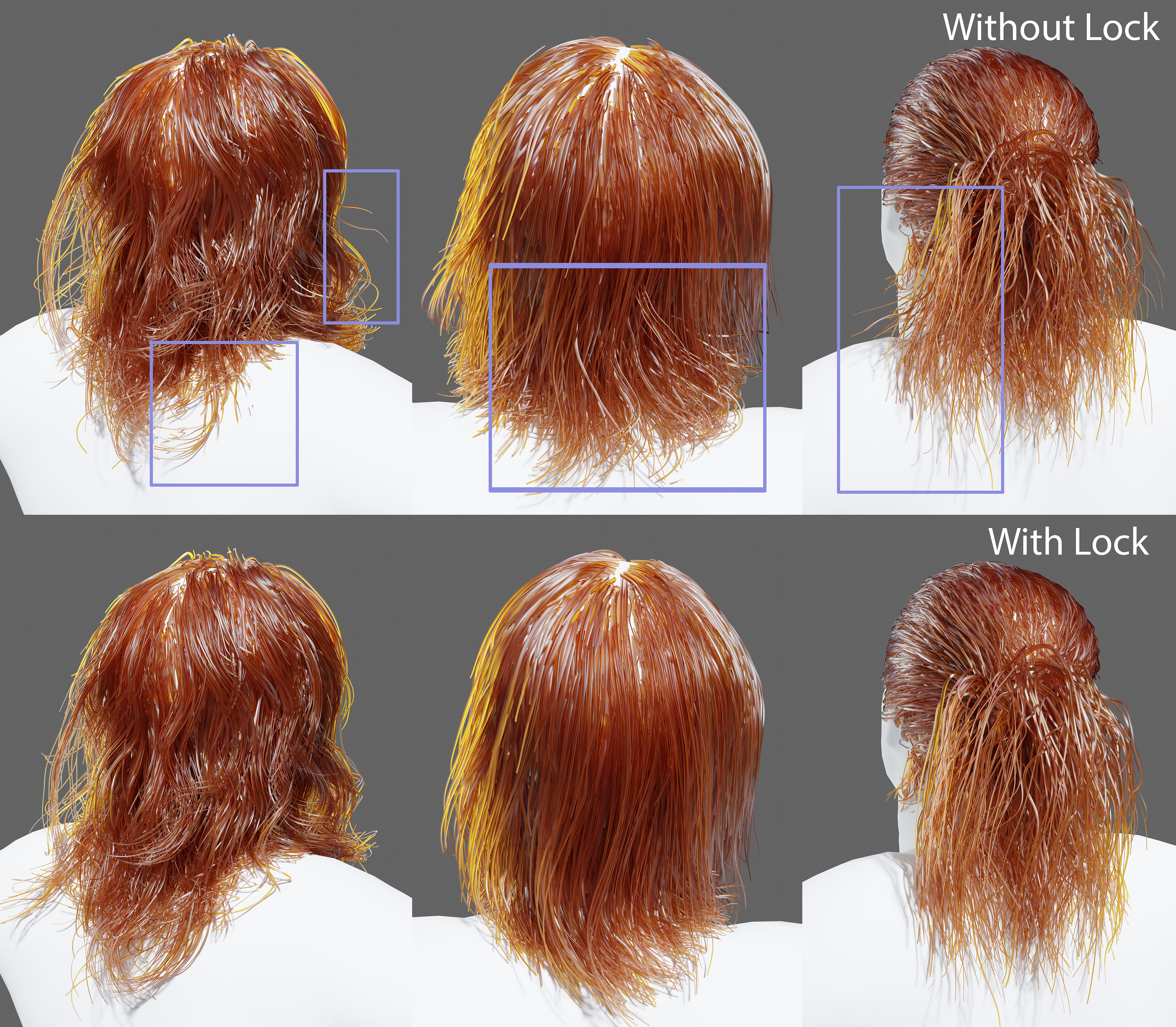}
    \caption{\textbf{Lock Based Dynamics} Our proposed neural lock effectively removes artifact due to similar rest strand code among multiple grooms for the neural simulator during training. Unexpected deformation in straight strands, expanding in ponytail and unresolved collision, can be observed without the local lock information.}
    \label{fig:lockVSNoLock}
\end{figure}

We augment our neural network input with the average strand latent code within a local neighborhood to encodes information to distinguish different configurations for a latent strand representation. This allows the neural simulator to compute the appropriate deformations. We demonstrate in Fig.~\ref{fig:lockVSNoLock} that this improves the quality of the produced results. Without such proximity-based information, we find that the simulator averages out simulation results when trained with multiple hair styles simultaneously, noticeably degrading quality of deformations.

\subsubsection*{Cosserat Energy Formulation}
\begin{figure}
    \centering
    \includegraphics[width=0.95\linewidth]{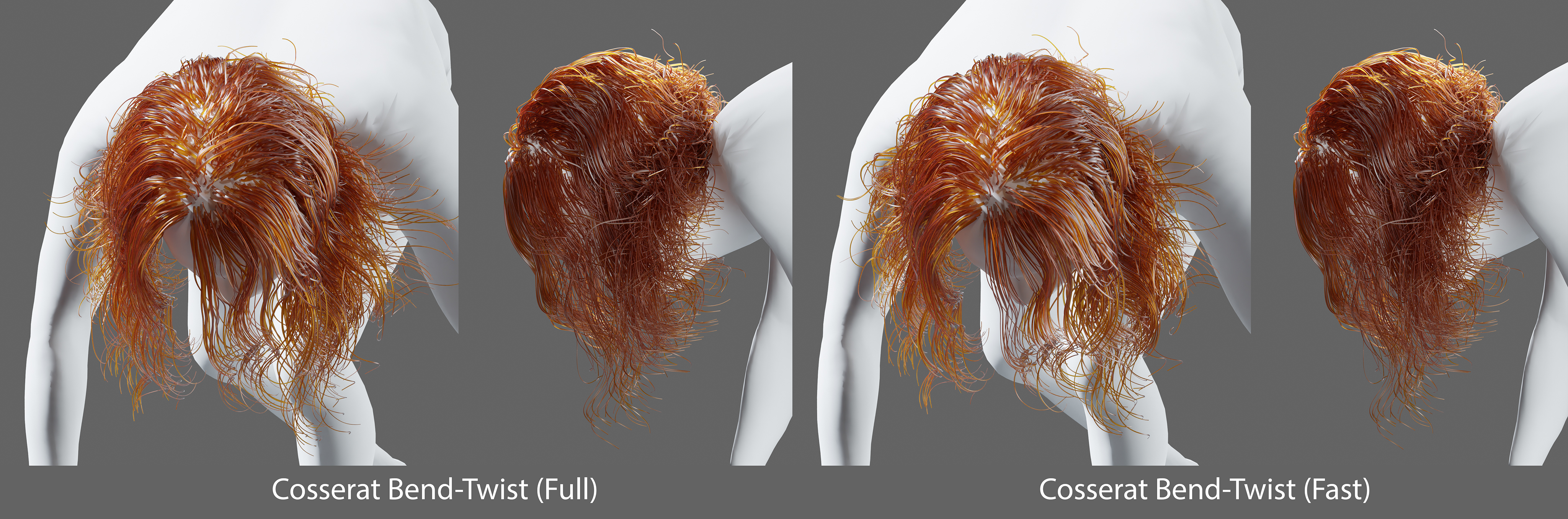}
    \caption{\textbf{Efficient Cosserat Formulation} We utilize fast parallel transport in our Cosserat energy formulation which significantly speeds up the computation while generating similar deformation compared to the full model.}
    \label{fig:cosseratSlowVSFast}
\end{figure}

We evaluate a Cosserat energy during training only, nonetheless high performance is critical to obtain fast training times. Fig.~\ref{fig:cosseratEnergyComparison} shows that our Cosserat energy formulation accurately reproduces the deformation of the strands under gravity. Our proposed fast parallel transport is 6 times faster at an average time of 0.16 seconds than the more accurate formulation (average time 0.98 s) and with similar efficiency as the energy proposed in Quaffure (average time 0.14) while producing similar strand deformation, see Fig.~\ref{fig:cosseratSlowVSFast}.

\subsection{Strand Generalization}

Our proposed method exhibits strand generalization capabilities. By leveraging a library of grooms, each possibly comprising over hundreds of thousands of strands, we train the network on a sparse subset of strands in the order of thousands. Notably, our approach enables the network to generate high-quality dynamic deformations of the full groom, thereby demonstrating its ability to generalize effectively to strands within a groom. This property facilitates efficient training and allows the trained network to simulate an arbitrary number of strands using UV coordinates and remaining strand encodings as input, scalable up to the full groom, as illustrated in Fig.~\ref{fig:neuralUpsampling}. This eliminates the need for any additional upsampling procedures. In contrast, prior methods often use an efficient 2D texture space representation for the guide strands~\cite{zhou2023groomgen, stuyck2024quaffure}, these methods require an additional upsampling operation~\cite{hsu2024real} to convert the sparse set of guide strands to a full groom, typically consisting of more than a hundred thousand strands.

\subsection{Mesh-based Rigged Hair}

\begin{figure}
    \centering
    \includegraphics[width=0.95\linewidth]{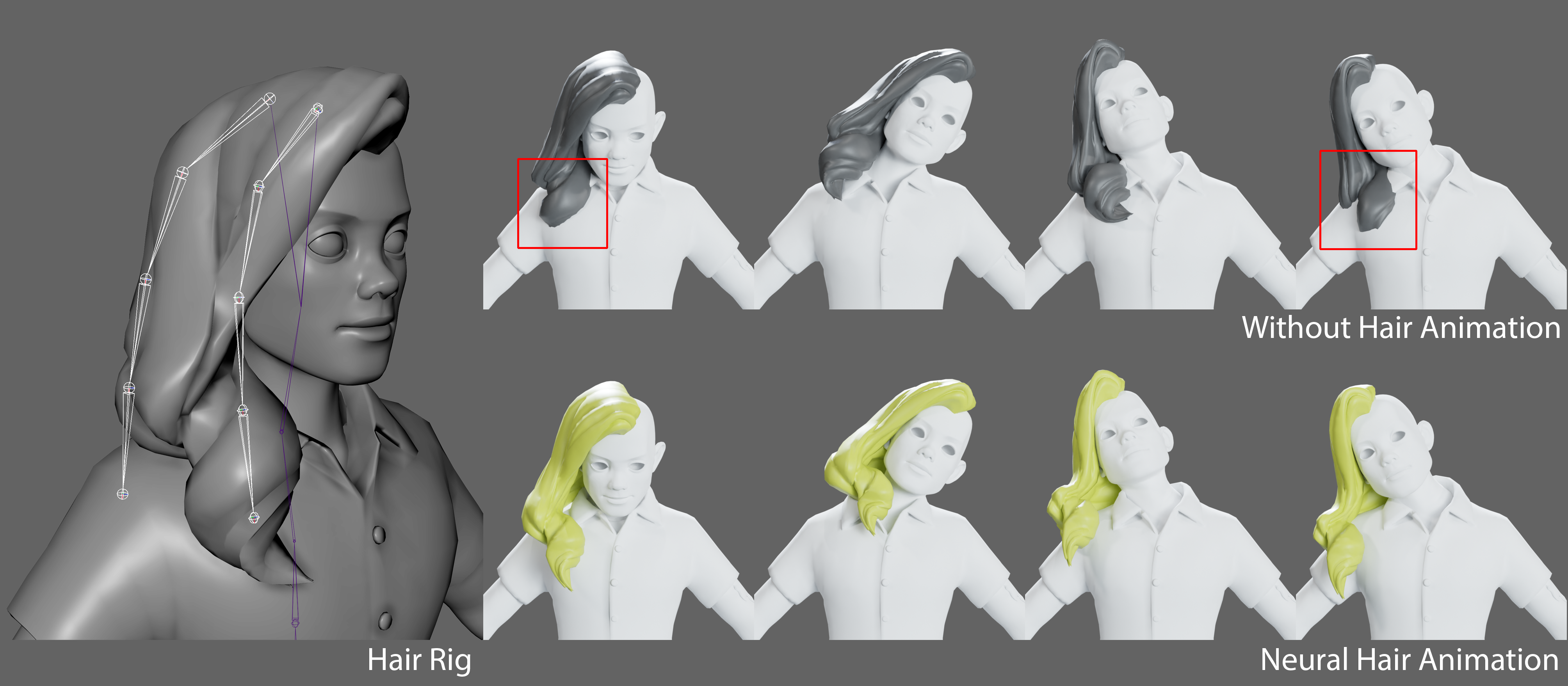}
    \caption{Predicting quasi-static bone animation using our neural hair simulation enables an ultra-low-compute solution for stylized avatar hair deformations which runs in sub-microseconds. Given neck joint rotations as input, our network outputs hair bone animation which maintains hairstyle while respecting gravity and avoiding intersections with the underlying body.}
    \label{fig:neuralMeshHair}
\end{figure}

As a practical demonstration of our contributions, we apply our method to the animation of rigged hair, which is deformed through a combination of bone animation and linear blend skinning of the mesh as shown in Fig.~\ref{fig:neuralMeshHair}. This enables automatic hair animation without artist intervention at a compute cost that allows inference for a large number of avatars simultaneously. To achieve this, we train a significantly smaller quasi-static version of our neural simulator by randomly sampling neck joint rotations, using these as input to the network. We modify the collision loss to minimize intersections of the mesh of the deformed hair against the body. The network then generates joint angles for the joint chains which are treated as hair strands to which the hair is rigged, effectively retargeting the head motion to the hair.

\subsection{Performance and Implementation Details}
\label{subsec:performance}

We implement our pipeline using PyTorch, and evaluate all methods on an Intel Core i7 CPU and one NVIDIA RTX 4080 Laptop GPU. Our neural simulator is implemented as a multi-layer perceptron which allows for easy implementation and support for runtime optimizations using libraries like ONNX. Our network consists of 2 layers with hidden dimension 256 with output dimension 72 and input dimension 748 in the dynamic model. In the case where predicting quasi-static deformation is sufficient, we can skip the inertia potential and replace the motion history with a single pose, resulting in a much smaller network input of size 100. For predicting a groom with 3,000 strands, our static network runs at 0.105 ms, while the dynamic counterpart runs at a 0.189 ms due to the larger network input size to capture body pose history. For a full groom consisting of 120,000 strands, our dynamic network runs at 6.872 ms. In comparison, Dynamic Quaffure has an inference time of 3.89 ms which is an order of magnitude slower. Our implementation of GroomGen reports a runtime cost of 5.364 ms for 3,000 strands.
For mesh-based rigged hair, we achieve high performance since we can significantly scale down the neural network size to only 2 layers with hidden dimension 16. Custom C++ inference code runs at a mere 0.3 microseconds using single-threaded execution on CPU.

\begin{table}[]
    \centering
    \begin{tabular}{l c c c c c}
         & XPBD CPU& XPBD GPU&  \multicolumn{3}{c}{Ours} \\ \toprule
        strand count & 3k & 3k & 3k & 50k & 120k \\ 
        avg. time (ms)& 7,200 & 140 & 0.189 & 2.99& 6.872\\ \bottomrule 
    \end{tabular}
    \caption{Comparison of performance per frame for generating strand deformation using an XPBD physics simulator and our neural model. Our lightweight network is orders of magnitude faster with plausible deformations.}
    \label{tab:runtime_performance_table}
\end{table}

\section{Limitations and Future Work}

Our method has the advantage of directly generating dynamic hair simulations from a sequence of boundary conditions, without requiring a time-evolution process. This leads to excellent stability, but at the cost of motion variety. Specifically, our method produces deterministic results, meaning that the same input sequence will always yield the same output, regardless of the preceding motion. In contrast, real-world hair behavior is influenced by its past configurations. Our current examples primarily demonstrate interactions and collisions with the underlying body, specifically focusing on areas frequently seen during training such as the head, neck, and upper torso. Although hair occasionally interacts with hands in our dataset, these instances are too sparse to enable the model to effectively respond to collisions with other parts of the body. To address this limitation, we plan to investigate interactions with external objects and the character's hands. Additionally, we intend to explore hair-garment interactions and increase body shape variability by moving beyond the statistical body model. Like prior work, our current model is trained with fixed material parameters. As future work, we plan to condition the neural simulator on the material parameters.

\section{Conclusion}

We present a neural simulator for dynamic hair simulations, marking the first time that self-supervised training has been successfully applied to achieve realistic and efficient hair dynamics. Our approach provides a crucial component for automated avatar reconstruction to enable efficiently simulated hair without manual intervention. We introduce a novel direct mapping between boundary condition history and strand deformations, significantly simplifying the training procedure and inference compared to prior work while maintaining dynamic results that react naturally to body movement. By operating at the strand level, our method achieves efficient runtime computations with limited memory requirements, outperforming state-of-the-art approaches. Through a variety of results, we demonstrate the robustness and generalizability of our approach, showcasing its ability to handle different grooms, body motion, and shape.


\bibliographystyle{ACM-Reference-Format}
\bibliography{biblio}

\clearpage
\input{figureOnlyPages}


\end{document}

%% file: commands.tex
\definecolor{peach}{rgb}{ 0.943, 0.188, 0.526}
\definecolor{plum}{rgb}{ 0.858, 0.188, 0.478}
\definecolor{muted_navy_blue}{RGB}{63, 75, 166}
\definecolor{muted_sky_blue}{RGB}{134,166,213}
\definecolor{federal_blue}{RGB}{0,96,240}
\definecolor{regulation_red}{RGB}{226, 20, 79}
\definecolor{federal_gold}{RGB}{240, 212, 14}

\usepackage{subcaption}
\usepackage{wrapfig}
\usepackage{enumitem}
\usepackage{graphicx}
\usepackage{amsmath}
\usepackage{booktabs}
\usepackage{enumitem}

\usepackage{bm}
\usepackage{amsmath}
\usepackage{colortbl}
\usepackage{multirow}
\usepackage{tikz}
\usepackage{pgfplots}
\usepackage{wrapfig}
\usepackage{cases}

\newcommand{\mytilde}{\raise.17ex\hbox{$\scriptstyle\mathtt{\sim}$}}
\usepackage{algpseudocode}
\usepackage[ruled,vlined]{algorithm2e} 

\usepackage{arydshln}

%% file: figureOnlyPages.tex
\begin{figure*}
    \centering
    \includegraphics[width=0.95\linewidth]{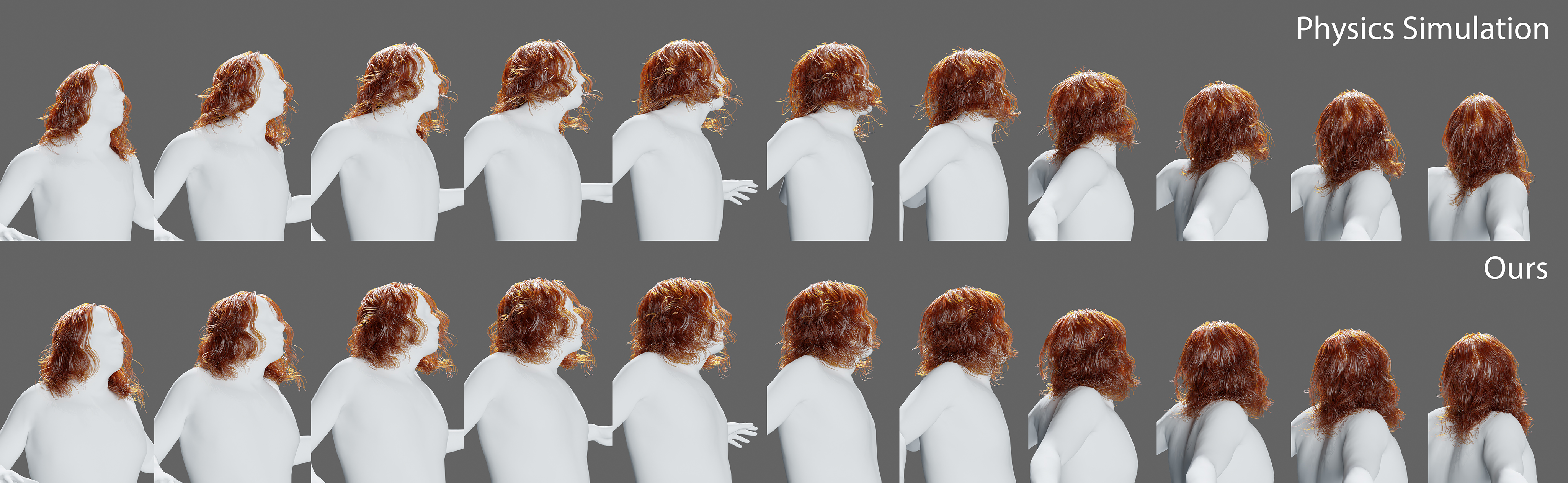}
    \caption{\textbf{Physics Simulation Comparison} We evaluate our neural simulation results against those of a widely adopted, highly efficient XPBD simulation model. While physics-based simulations offer the highest quality results with dynamic behavior, they come at a significant computational cost. In contrast, our neural method captures the dynamic behavior of hair, albeit with more damping, but achieves this at a fraction of the computational cost.}
    \label{fig:simVSNeuralocks}
\end{figure*}

\begin{figure*}
    \centering
    \includegraphics[width=0.95\linewidth]{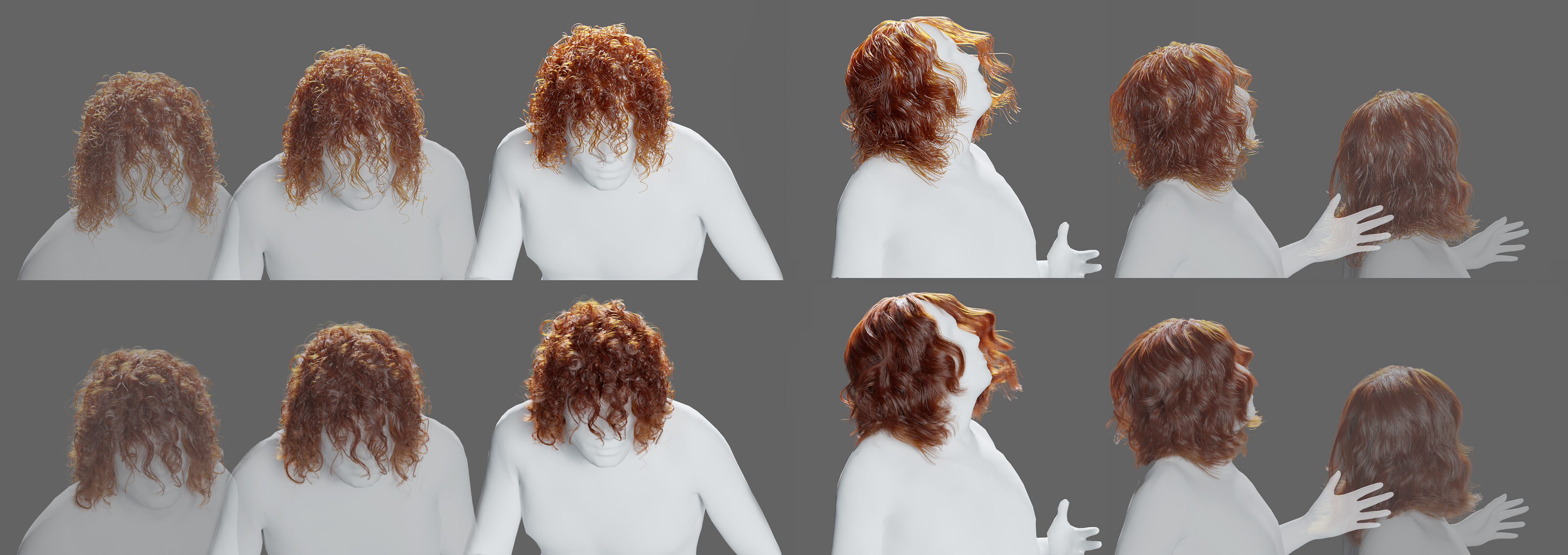}
    \caption{\textbf{Strand Generalization} Our neural simulator, efficiently trained on a sparse subset of 3,000 strands (top), showcases its robust generalization capabilities by predicting the deformation of all remaining strands in the full groom, comprising of a total of 120,000 strands (bottom).}
    \label{fig:neuralUpsampling}
\end{figure*}

\begin{figure*}
    \centering
    \includegraphics[width=0.5\linewidth]{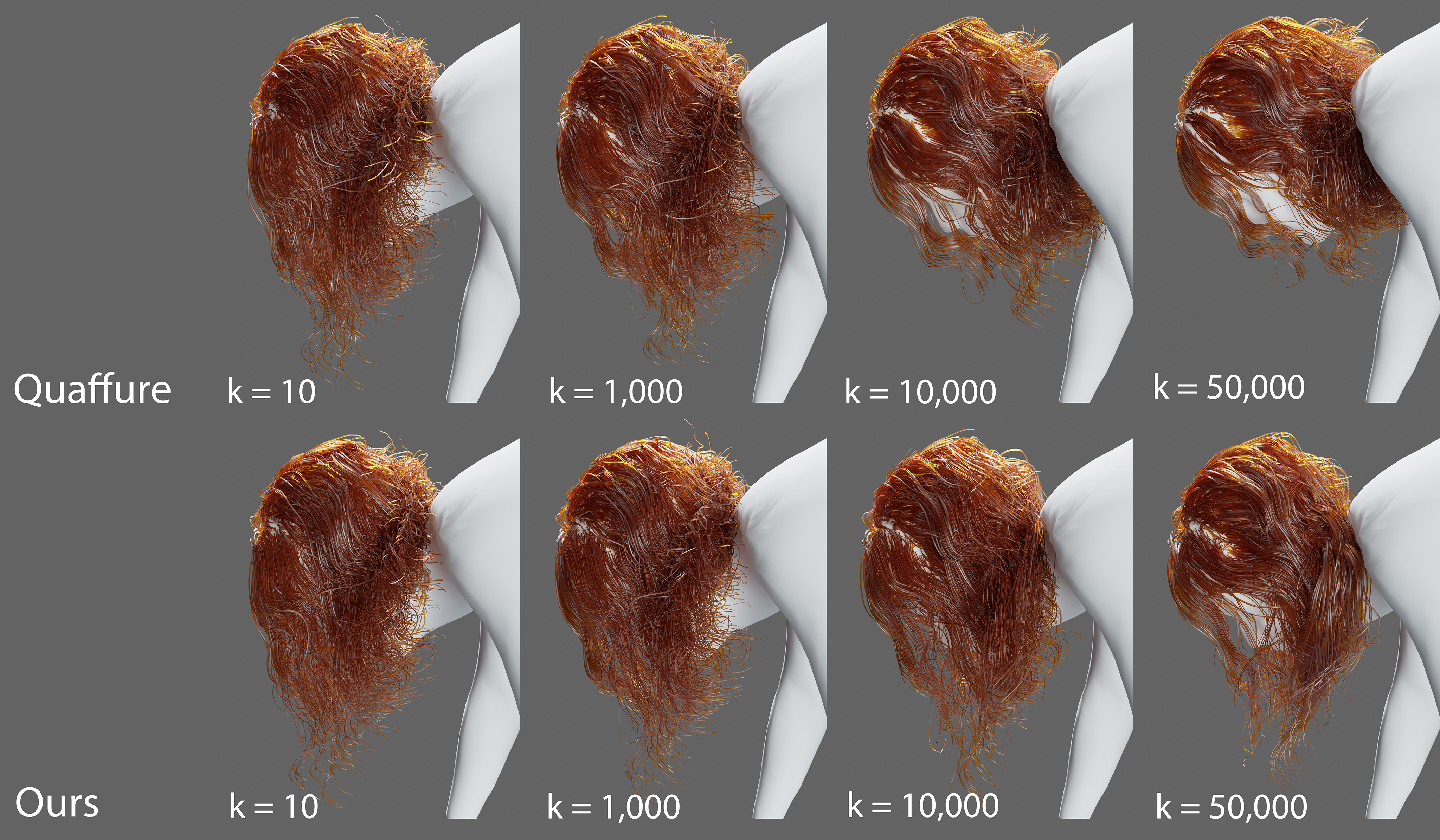}
    \caption{\textbf{Cosserat Energy for varying Stiffness} Our Cosserat energy formulation produces plausible deformation of the wavy strands when stiffness is increased, while the formulation presented in Quaffure produces overly stiff deformation which results in rigid deformations that fail to adhere to the gravity direction. In contrast, our formulation naturally twists the strands to reacts to gravity while accurately maintaining the strand rest shape.}
    \label{fig:cosseratEnergyComparison}
\end{figure*}

\begin{figure*}
    \centering
    \includegraphics[width=0.95\linewidth]{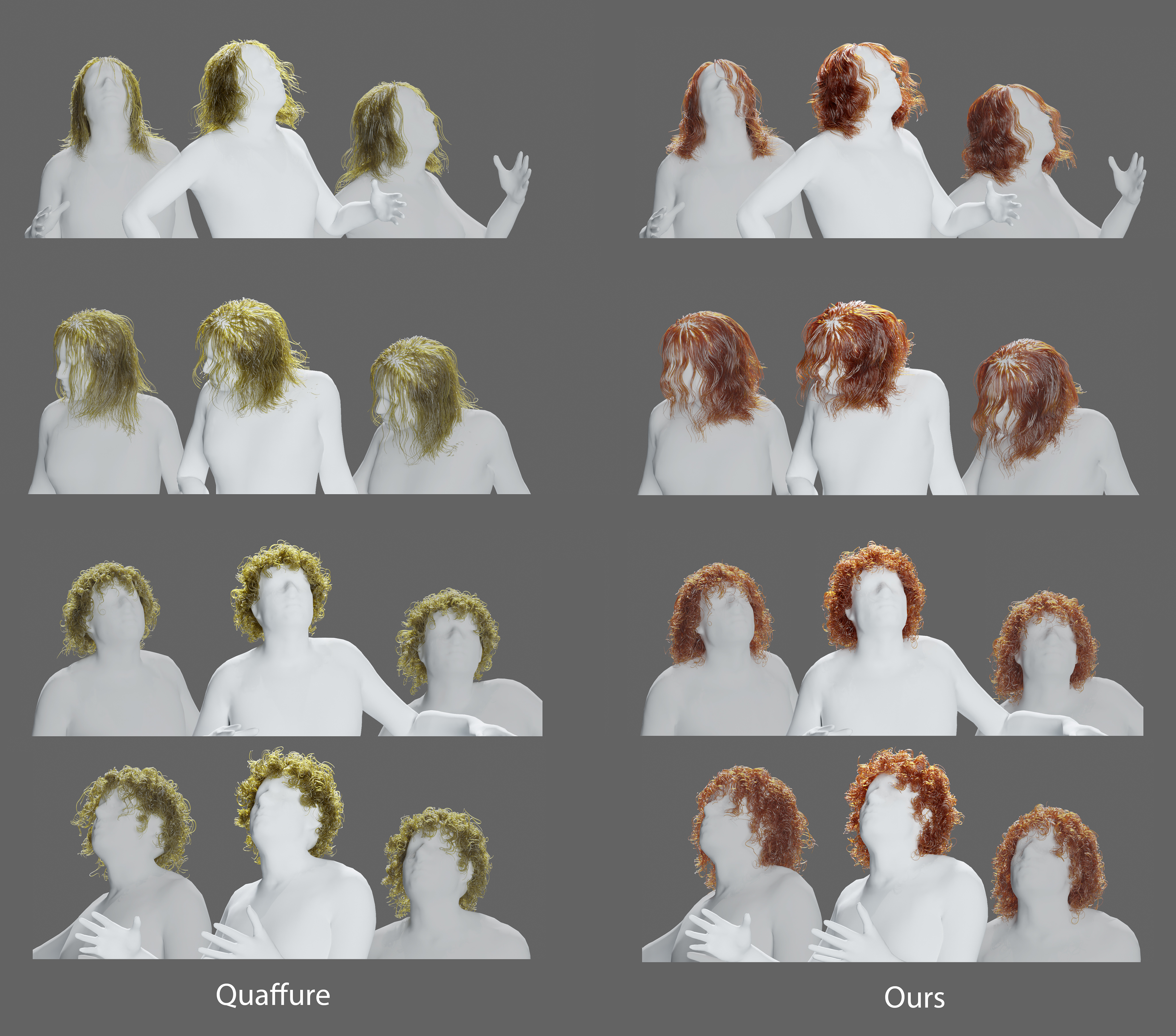}
    \caption{\textbf{Dynamic Quaffure Comparison} To ensure a fair comparison, we have extended Quaffure to incorporate dynamic behavior following our listed contribution and evaluated it against our approach. The results show that both methods produce comparable results bur ours does so at a significantly lower memory and performance footprint. While Quaffure can predict the full groom in a single forward pass, this comes at the cost of limited scalability, with a maximum strand count of 4096 in their implementation. Increasing strand counts requires larger maps, leading to further memory and runtime increases. In contrast, our method naturally scales to accommodate large strand counts. Furthermore, the Dynamic Quaffure model is significantly larger, weighing in at 692MB, whereas our model is much more compact at 1MB.}
    \label{fig:quaffureComparison}
\end{figure*}

\begin{figure*}
    \centering
    \includegraphics[width=0.95\linewidth]{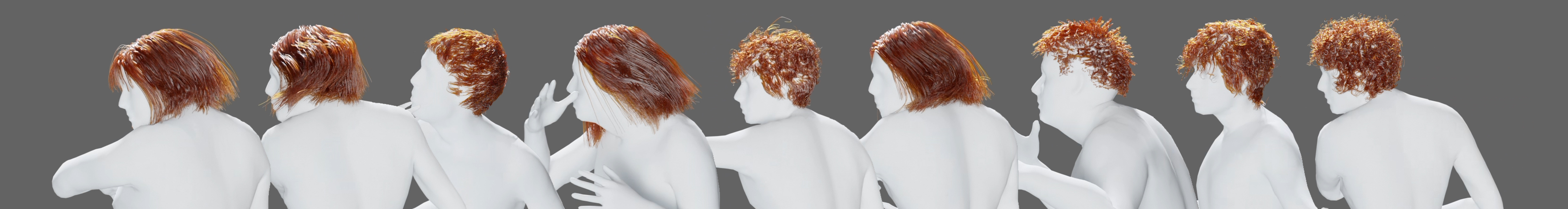}
    \caption{\textbf{Unseen Grooms} Our neural simulator generalizes to similar grooms that are
not in the training set, producing plausible strand deformation.}
    \label{fig:unseenGrooms}
\end{figure*}